\documentclass[a4paper,12pt,twoside]{article}
\usepackage{graphicx}
\usepackage{amsmath,amsfonts,amssymb,amsthm}
\newcommand{\Ii}[2]{{}^{#1}_{\phantom{#1}\!#2}}
\newcommand{\bR}{{\mathbb R}}

\newcommand{\Cset}{{\mathbb C}}
\newcommand{\Pset}{{\mathbb P}}
\newcommand{\be}{\begin{equation}}
\newcommand{\ee}{\end{equation}}
\newcommand{\n}{{\bf n}}
\newcommand{\rr}{{\bf r}}

\newcommand{\go}{\mathfrak}                 
\newcommand{\ga}{\go A}
\newcommand{\gH}{\go H}
\newcommand{\aaa}{A_\alpha}
\newcommand{\gba}{g_{\beta\alpha}}
\newcommand{\ha}{H_\alpha}
\newcommand{\la}{\Lambda_\alpha}
\newcommand{\gab}{g_{\alpha\beta}}
\newcommand{\gHa}{{\go H}_\alpha}

\def\slashiii#1{\setbox0=\hbox{$#1$}#1\hskip-\wd0\hbox to\wd0{\hss\sl/\/\hss}}
\DeclareGraphicsExtensions{.jpg,.pdf,.gif,.png,.eps}
\begin{document}
\newtheorem{theorem}{Theorem}
\title{On Quantum Iterated Function Systems}
\author{A. Jadczyk\\
Institute of Theoretical Physics, University of
Wroc{\l}aw,
\\Pl. Maxa Borna 9, 50 204 Wroc{\l}aw, Poland}


\maketitle
\begin{abstract}Quantum
Iterated Function System on a complex projective space is defined
by a family of linear operators on a complex Hilbert space. The
operators define both the maps and their probabilities by one
algebraic formula. Examples with conformal maps (relativistic
boosts) on the Bloch sphere are discussed.\end{abstract}
\section{Introduction}
Iterated Function Systems \cite{barnsley}  generate fractal sets
due to non-commutativity of maps. In quantum theory position and
momentum operators do not commute (which leads to the
Heisenberg's uncertainty relations), and also different components
of spin do not commute. This suggests that fractal patterns and
chaos may arise as a result of certain quantum measurement
processes, for instance in a continuous monitoring of several
noncommuting observables. A typical continuous monitoring takes
place, for instant, in a cloud chamber. Different regions of the
chamber are active in parallel, and they are activated
sequentially, each at a different time, by a charged quantum
particle that leaves the track. Moreover, in Heisenberg's picture,
position operators at different time do not commute. Parallel
arrangement of an experimental setup is realized by the addition
of operators, while serial arrangements lead to multiplication of
the corresponding operators \cite{skala}. While sums of operators
are commutative, their products, in general, depend on the order
of factors. Sums of terms usually appear in time evolution
generators. Products appear in transitions from one quantum state
to another, as the result of the wave function collapse resulting
from measurement events. Repeated application, with non-commuting
operators, leads to iterated function systems, chaos and fractal
attractors on the manifold of quantum states. In the present paper
we will describe the present status of this new research
avenue, and we point at some open questions.

\section{The two--sphere $S^2$ as the canvas}
The flagship example of a classical "Iterated Function System" (in
short: IFS) is the Sierpinski fractal.
\begin{figure}[!htb]
  \begin{center}
    \leavevmode
      \includegraphics[width=5cm, keepaspectratio=true]{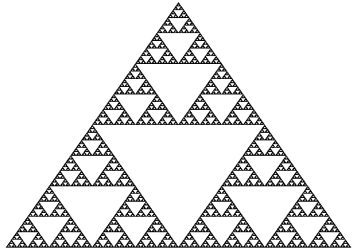}
  \end{center}
\caption{The classical fractal: Sierpinski Triangle generated by an
Iterated Function System.}\label{sierpinski}
\end{figure}
It is generated by an application of $3\times 3$ matrices $A[i],
 i=1,2,3,$ in a random order, to the vector:
\begin{equation}
v_0=\begin{pmatrix}x_0\\ y_0\\ 1\end{pmatrix}
\end{equation}
where $A[i]$ is given by
\begin{equation}
A[i]=\begin{pmatrix}0.5&0&ax_i\cr 0&0.5&ay_i\cr0&0&1\end{pmatrix}
\end{equation}
and $ax_1=0.0, ay_1=0.0, ax_2=0.5, ay_2=0.0, ax_3=0.25, ay_3=0.5.$
(Our $3\times 3$ matrices encode  affine transformations (of a
two--plane) -- usually separated into a $2\times 2$ matrix and a
translation vector.) At each step one of the three transformations
$A[i], i=1,2,3$ is selected with probability $p[i]=1/3$. After
each transformation the transformed vector is plotted on the
$(x,y)$ plane.

The important property of the maps $A[i]$ is that they are
contractions. Sierpinski triangle, as well as another well known
example, the fern \cite{barnsley}, live on a 2-dimensional plane.
Quantum iterated function systems (QIFS) live on complex
projective spaces, the simplest one being $CP(1)$ - a
2-dimensional sphere $S^2,$  also known as the Bloch sphere.

In fact, there are at least two ways in which $S^2$ is important
in physics. First, as the Bloch sphere, it represents pure states
of the simplest quantum system - spin $1/2.$ Second, it represents
directions in our three-dimensional space. The last statement is
not relativistically invariant. But there is another,
relativistically invariant interpretation of $S^2,$ namely as the
space of directions of light rays. We will start with this second
interpretation.

\subsection{$S^2$ as the projective light cone}
Affine transformations form a natural group of transformations
acting on the plane. What is the natural group of transformations
acting on the two-sphere? One would think it is the rotation group
$O(3)$. But rotations are volume preserving and they would not
mimics contractions. The next candidate in line is the Lorentz
group $O(3,1).$ It is not so well known that the Lorentz group
acts on a manifold that is diffeomorphic to the sphere $S^2$ in a
natural way. One way to see that this is the case is to notice
that the Lorentz group is the group preserving the space-time
metric $s^2 = -x_0^2+ x_1^2+x_2^2+x_3^2$ and thus the light cone
$C=\{x=(x^0,x^1,x^2,x^3): -(x^0)^2+(x^1)^2+(x^2)^2+(x^3)^2=0\},$
and therefore, because it acts by linear transformations, also the
projective light cone $\Pset C,$ that is the set of equivalence
classes ${\hat x}:x\in C,$ with respect to the equivalence
relation $R\ \subset (C\setminus\{0\})\times (C\setminus\{0\})$
where $xRy\ \mbox{iff}\ x=\lambda y ,\lambda\neq 0 .$ Each
equivalence class has a unique representative with $x^0=1,$ so
$\Pset C$ can be identified with the sphere $S^2=\{{\n}\in\bR^3:
\n^2=1\}.$ The Lorentz group $O(3,1)$ consists of $4\times4$ real
matrices $\Lambda=(\Lambda\Ii{\mu}{\nu})$ satisfying $\Lambda^T
\eta \Lambda =\eta,$ where $\eta=(\eta_{\mu\nu})=diag (-1,1,1,1)$
is the diagonal metric matrix. The action $ S^2\ni\n\mapsto
\Lambda(\n)$ of $O(3,1)$ on $S^2$ is given explicitly by the
formula: \be
\Lambda(\n)^i=\frac{\Lambda\Ii{i}{0}+\Lambda\Ii{i}{j}n^j}{\Lambda\Ii{0}{0}+\Lambda\Ii{0}{j}n^j},
\label{eq:lc}\ee (we use Einstein's summation convention over
repeated indices). The group $O(3,1)$ has four connected
components. We will need only the connected component of the
identity $SO_{+}(3,1)$ consisting of those matrices $\Lambda$ in
$O(3,1)$ for which $\det(\Lambda)=1$ and $\Lambda\Ii{0}{0}>0.$ The
group $SO_{+}(3,1),$ though connected, is not simply connected.
Its simply connected double covering group is the group
$SL(2,\Cset)$ of $2\times2$ complex matrices of determinant $1.$
By polar decomposition every matrix $A\in SL(2,\Cset)$ can be
uniquely decomposed $A=PU$ into a positive part $P$ and a unitary
part $U\in SU(2).$\footnote{In Relativity the positive matrices
represent ``Lorentz boosts."} The group $SU(2)$ is the double
covering of the rotation group $SO(3).$ Nontrivial positive
matrices in $SL(2,\Cset)$ have two eigenvalues $\lambda_1<1$ and
$\lambda_2=1/\lambda_1>1.$ It is the positive matrices in
$SL(2,\Cset )$ that will generate our iterated function systems.

To describe the $2:1$ group homomorphism $A\mapsto\Lambda(A)$ from
$SL(2,\Cset )$ to $SO_{+}(3,1),$ and also to describe
algebraically the action of $SL(2,\Cset )$ on $S^2$ it is
convenient to use the Pauli spin matrices
$\sigma_0,\,\sigma_1,\,\sigma_2,\,\sigma_3$ defined by
$$\sigma_0\:=I\:=\:\left(\begin{array}{cc}1&0\\0&1\end{array}\right)\quad\sigma_1\:=\:\left(\begin{array}{cc}
0& 1\\1& 0 \end{array}\right)\quad
\sigma_2\:=\:\left(\begin{array}{cc} 0& -i\\i& 0
\end{array}\right)\quad \sigma_3\:=\:\left(\begin{array}{cc} 1&
0\\0& -1 \end{array}\right).$$  The homomorphism $SL(2,\Cset
)\rightarrow SO_{+}(3,1)$ is then given by the formula: \be
\Lambda(A)\Ii{\mu}{\nu}=\frac{1}{2}\mbox{Tr} (\sigma_\mu A
\sigma_\nu A^\star), \ee where $A^\star$ denotes the Hermitian
conjugate of $A.$ Every Hermitian $2\times 2$ matrix $X$ can be
uniquely represented as $X=x^\mu\sigma_\mu$, with $x^\mu$ real.
For every $\epsilon\in [0,1],$ and every unit length vector $\n\in
S^2$ let \be P(\n,\epsilon)=\frac{1}{2}(I+\epsilon \sigma(\n)),\ee
where $\sigma(\n )\doteq n^1\sigma_1+n^2\sigma_2+n^3\sigma^3.$  It
is easy to see that a Hermitian matrix $X\neq I$ is an idempotent
if and only if it is of the form $X=P(\n,1)$ for some $\n\in S^2.$
We will write $P(\n )\doteq P(\n,1).$ It is also easy to check
that a matrix $P$ is positive if and only if it is of the form
$P=c\ P(\n,\epsilon)$, for some $c>0,\ \epsilon\in [0,1],\ \n\in
S^2.$ Notice that $\det(P)=1$ if and only if $\epsilon<1$ and
$c=2(1-\epsilon^2)^{-1/2}.$ We will use the matrices
$P(\n,\epsilon),$ with the same $\epsilon$ but different vectors
$\n$ to generate IFS-s on $S^2.$ The formula (\ref{eq:lc})
describing the action of the Lorentz group on $S^2$ is not the
most convenient one for our needs. Another way of describing the
same action is by noticing that, for $\rr\in S^2,$ the following
identity holds\footnote{A more general formula is discussed in
Sec. \ref{sec:ds}, Eq. (\ref{eq:pgba})} \be P(\n ,\epsilon )P(\rr
)P(\n ,\epsilon )=\lambda(\epsilon,\n,\rr) P(\rr^\prime),
\label{eq:lambda1} \ee where $\lambda(\epsilon,\n,\rr)\geq 0$ is
given by: \be
\lambda(\epsilon,\n,\rr)=\frac{1+\epsilon^2+2\epsilon (\n\cdot\rr
)}{4} \label{eq:lambda2}, \ee while  \be
S^2\ni\rr^\prime=\frac{(1-\epsilon^2)\rr+2\epsilon(1+\epsilon(\n\cdot\rr
))\n}{1+\epsilon^2+2\epsilon (\n\cdot\rr )} \ee where $(\n\cdot\rr
)$ denotes the scalar product \be \n\cdot\rr=n_1r_1+n_2 r_2+n_3
r_3.\label{eq:rprime}\ee \footnote{Because of the property
$\lambda(\epsilon,\n,\rr)\geq 0$, $\lambda$ will be later on --
cf. Eq. (\ref{eq:pi}) and Sec. \ref{sec:ds} --  interpreted as the
probability of a jump. This interpretation is natural within the
quantum measurement theory, but its physical meaning within the
framework of light directions and Lorentz boosts is not clear.}
The map $\rr\mapsto\rr^\prime$ is the same as the one described in
Eq. (\ref{eq:lc}), with
$\Lambda=\Lambda\left(2P(\n,\epsilon)/\sqrt{1-\epsilon^2}\right).$
Notice that the dilation coefficient $2/\sqrt{1-\epsilon^2}$ is
not important here, because it would cancel out anyway in
Eq.(\ref{eq:lc}). The transformation $x^\mu\mapsto
\Lambda\Ii{\mu}{\nu}x^\nu$ implemented by $\Lambda$ can be
explicitly described by the formula known from texts on special
relativity: \begin{eqnarray} x^{0\prime}&=&\cosh (\alpha)
x^0+\sinh(\alpha)({\bf x}\cdot\n ),\\\nonumber {\bf
x}^\prime&=&{\bf x}-({\bf
x}\cdot\n)\n+[\sinh(\alpha)x^0+\cosh(\alpha)({\bf x}\cdot\n)]\n ,
\end{eqnarray} where the "velocity" $\beta=\tanh
(\alpha)=2\epsilon/(1+\epsilon^2).$\footnote{Notice that for
$\epsilon\rightarrow 1,$ $\beta\rightarrow 1$ - the velocity of
light. In this limit the maps $\rr\mapsto\rr^\prime$ degenerate to
$\rr\mapsto\n$ and become non-invertible.} What is important for
us, is the fact that the coefficient $\lambda(\epsilon,\n,\rr)$ in
Eq.(\ref{eq:lambda1}) is positive, and thus can be interpreted as
a (relative) probability associated with the transformation
$\rr\mapsto\rr^\prime .$ In other words: relative probabilities
associated to maps implemented by $P(\n,\epsilon )$ are naturally
associated with the maps. It should be noticed that positivity of
$\lambda$ is guaranteed by the algebraic properties of the
operators involved. Indeed, because $P(\n ,\epsilon)=P(\n
,\epsilon )^\star ,$ and because $P(\rr)=P(\rr)^\star=P(\rr)^2$ is
an orthogonal projection, the right hand side in
Eq.(\ref{eq:lambda1}) can be represented as $A^\star A$, with
$A=P(\rr )P(\n,\epsilon)$, and is, therefore, automatically
positive. \section{Quantum Iterated Function Systems on $S^2.$}
Given a sequence $\n_i,\, i=1,\ \ldots\ ,N$ of vectors in $S^2$ we
associate with this sequence an iterated function system
$\{w_i,p_i\}$ on $S^2$, with place dependent probabilities,
defined as follows: \be w_i(\rr )=\rr^\prime=
\frac{(1-\epsilon^2)\rr+2\epsilon(1+\epsilon(\n_i\cdot\rr
))\n_i}{1+\epsilon^2+2\epsilon (\n_i\cdot\rr )}, \ee \be p_i(\rr
)=\frac{\lambda(\epsilon,\n_i,\rr)}{\sum_{j=1}^N
\lambda(\epsilon,\n_j,\rr)}.\label{eq:pi} \ee System $\{w_i,p_i\}$
defined by these formulae will be called a {\it Quantum Iterated
Function Systems}\ or {\it QIFS}\ . \footnote{A short
justification for the term ``quantum" will be given in the closing
section of this paper.} The formula for the probabilities
simplifies whenever \be\sum_{i=1}^N \n_i=0.\ee In the following we
will always assume that the vectors $\n_i$ defining the
transformations $w_i$ add to zero. In this case $p_i$ are given
by: \be p_i(\rr )=\frac{1+\epsilon^2+2\epsilon (\n_i\cdot\rr
)}{N(1+\epsilon^2)}. \ee In \cite{jadob02} we examined QIFS
corresponding to several most symmetric configurations, where the
vectors $\n_i$ were placed at the vertices of regular polyhedra:
tetrahedron (4), octahedron (6), cube (8), icosahedron (12),
dodecahedron (20), double tetrahedron (8), icosidodecahedron (30).
In each case numerical simulation of the Markov process, starting
with a random original point, lead to fractal-like patterns on the
sphere. For $\epsilon$ close to $1$ the operators $P(\n,\epsilon)$
are close to projections, therefore the attraction centers are
very distinctive. For $\epsilon$ close to $0$ the operators
$P(\n,\epsilon)$ induce maps close to the identity map - the
patterns are fuzzy. Typical patterns are shown in Fig. 1.
\begin{figure}[ht] \centering \includegraphics[width=10cm,
keepaspectratio=true]{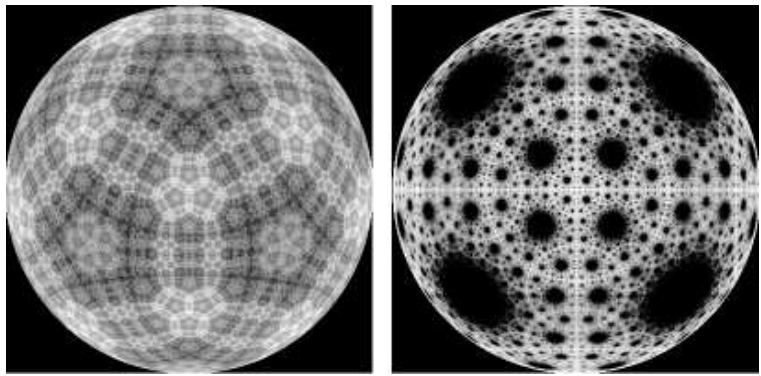}\caption{Quantum Dodecahedron
($\epsilon=0.78$) and Quantum Octahedron ($\epsilon=0.58$) The
darker the place, the smaller probability of it being visited. }
\end{figure}

It seems that the fractal dimension depends on the value of $\epsilon$. The Hausdorff dimension of the limit
set, for the tetrahedral case, has been numerically estimated in
Ref. \cite{gjast} and shown to decrease from 1.44 to 0.49 while
$\epsilon$ increases from 0.75 to 0.95.
\section{Transfer (Markov) Operator and Invariant Measure} 
\begin{figure}[ht] \centering \includegraphics[width=11cm,
keepaspectratio=true]{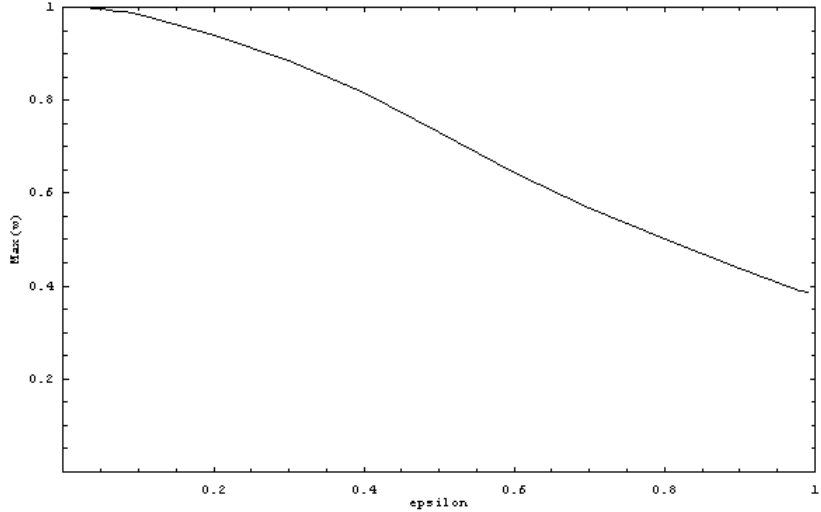}\caption{$\epsilon$--dependence of
the average contraction parameter $Max(w)$ for the quantum
Octahedron}\label{fig:max}
\end{figure}

Stenflo
\cite{stenflo02} gives a useful, brief review of the problem of
the existence and uniqueness of the invariant measures, which is
quite useful in our case. We will follow the notation and the
terminology of \cite{stenflo02}. The transfer operator $T$ for the
system is defined by the formula: \be (Tf)(\rr ) = \sum_{i=1}^N
p_i(\rr) f(w_i(\rr )), \ee where $f\in C(S^2)$ - the space of all
continuous functions on $S^2.$ By the Riesz representation theorem
$T$ induces the dual operator $T^\star :\mu\mapsto T^\star\mu$ on
the space $M(S^2)$ of Borel probability measures on $S^2$ via the
formula: $$\int_{S^2} (Tf)(\rr )\, d\mu (\rr ) = \int_{S^2}f(\rr
)\, d(T^\star \mu )(\rr ). $$ Since $S^2$ is compact, there always
exists an invariant probability measure $\mu$ that is invariant,
i.e. $T^\star \mu =\mu .$ Numerical simulations of QIFS seem to
indicate that such a measure is also unique, and that it is
concentrated on a unique attractor set, though different for
different $\epsilon\in(0,1).$ As each of the normalized operators
$2P(\n_i,\epsilon )/\sqrt{1-\epsilon^2}\in SL(2,\Cset)$ has two
eigenvalues, $(1+\epsilon)/(1-\epsilon)$, and
$(1-\epsilon)/(1+\epsilon) ,$ - the standard contraction arguments
do not apply. Nevertheless we have the following theorem:
\begin{theorem} For the Quantum Octahedron the invariant measure is unique
in the whole range of parameter $0<\epsilon<1.$\end{theorem} {\bf
Proof} In Ref. \cite{stenflo02} Stenflo states the following
classical results, attributed to Barnsley et al.
\cite{barnsley88}: {\it Let $\{(X,d),p_i(x),w_i(x),i\in
S=\{1,2,\ldots ,N\}\}$ be an IFS with place-dependent
probabilities, with all $w_i$ being Lipshitz continuous, and all
$p_i$ being Dini-continuous, and bounded away from $0.$ Suppose
\be \sup_{x\neq y}\sum_{i=1}^N p_i(x)\log
\left(\frac{d\left(w_i(x),w_i(y)\right)}{d(x,y)}\right)<0.\label{eq:in}\ee
Then the generated Markov chain has a unique invariant probability
measure.}\\ The log-average contraction condition (\ref{eq:in}) is
somewhat more general than the average contraction condition \be
Max(w)\doteq\sup_{x\neq y}\sum_{i=1}^N p_i(x)
\frac{d\left(w_i(x),w_i(y)\right)}{d(x,y)}<1.\label{eq:in1}\ee In
our case $w_i(x)$ and $p_i(x)$ are analytic, with $p_i(\rr
)\geq\frac{1-\epsilon^2}{N(1+\epsilon^2)}$ We made a numerical
estimation of the LHS of the inequality (\ref{eq:in1}), with $d$
being the natural, rotation invariant, $arc$ distance on $S^2,$
for the Quantum Octahedron, with $0<\epsilon<1$, and obtained the
epsilon-dependence showed in Fig. \ref{fig:max}
, thus assuring the uniqueness of the
invariant measure in this particular case.$\Box$
\smallskip

\begin{figure}[ht] \centering
\includegraphics[width=10cm,
keepaspectratio=true]{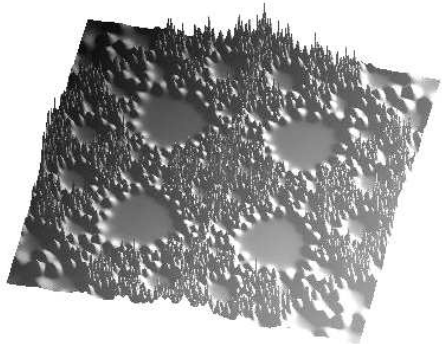}\caption{An
approximation of the invariant measure: Plot of $f_5(\rr)$ for
Quantum Octahedron ($\epsilon=0.58$)}\label{fig:markov5}
\end{figure}

Let $\mu_0$ be the natural, rotation-invariant, normalized measure
on $S^2.$ Then, for any finite $n$, the measure ${T^\star}^n\mu_0$
is continuous with respect to $\mu_0$ and therefore can be written
as
$${T^\star}^n\mu_0 (\rr)=f_n(\rr)\mu_0(\rr).$$ The sequence of
functions $f_n(\rr )$ gives a convenient graphic representation of
the limit invariant measure. In our case, the functions $f_n$ can
be computed explicitly via the following recurrence formula: \be
f_{n+1}(\rr )= \sum_{i=1}^{n}p_i\left(w_i^{-1}(\rr
)\right)\frac{d\mu_0\left(w_i^{-1}(\rr)\right)}{d\mu_0(\rr
)}f_n\left(w_i^{-1}(\rr)\right) \ee or, explicitly: \be
f_{n+1}(\rr)=\frac{(1-\epsilon^2)^4}{N(1+\epsilon^2)}\sum_{i=1}^N
\frac{f_n\left(w_i^{-1}(\rr)\right)}{(1+\epsilon^2-2\epsilon
\n_i\cdot\rr)^3} \ee where \be
w_i^{-1}(\rr)=\frac{(1-\epsilon^2)\rr-2\epsilon (1-\epsilon
\n_i\cdot\rr)\n_i}{1+\epsilon^2-2\epsilon\n_i\cdot\rr } \ee Fig.
(\ref{fig:markov5}) shows a plot of $\log(f_5(\rr)+1)$ for Quantum
Octahedron, $\epsilon=0.58,$ using the stereographic projection
$\n\mapsto z=\frac{n^1-i n^2}{1-n^3}$ from $S^2$ to the complex
plane. It should be noticed that via the stereographic projection
the maps $\rr\mapsto w_i(\rr)$ become fractional, and thus
conformal,  transformations of the complex plane: $z\mapsto
w_i(z)=\frac{az+b}{cz+d},$ with $a=1+\epsilon n_i^3,\,
b=\epsilon(n_i^1-\mbox{i} n_i^2),\, c=\epsilon(n_i^1+\mbox{i}
n_i^2),\, d=1-\epsilon n_i^3.$ 
\section{Concluding Remarks} In this section we will place QIFS
within a larger field of piecewise-deterministic Markov processes
and their connection to dissipative dynamics of mixed
quanto-classical dynamical systems. \subsection{Classical
dynamics} Usually classical dynamics is described by a 1-parameter
group $\phi_t$ of diffeomorphisms of a smooth manifold $X.$ In
classical mechanics $X$ is a symplectic manifold, the ``phase
space" of the system, and the flow $\phi_t$ is generated by a
Hamiltonian vector field on $X$. States of the system are simply
points of $X,$ statistical states are probabilistic measures on
$X.$ The set of all statistical states is convex, its extremal
elements are called pure states. These are Dirac measures -
concentrated at points of $X.$  The flow $\phi_t$ on $X$ gives
rise to a flow on the space of ``observables", that is functions
on $M,$ and to a flow on the space ``statistical states", that is
on the space $M(X)$ of probabilistic measures on $X$. If $X$ is
discrete, then we can't have a continuous flow on $X$, but we can
still have a continuous family of transformations acting on
observables and on statistical states. \subsection{Quantum
dynamics} Quantum theory is usually formulated in terms of linear
operators acting on a separable complex Hilbert space $\gH.$
Observables are represented by Hermitian elements of the algebra
$\ga=L(\gH )$ of all bounded linear operators on $\gH.$
Statistical states are positive, normalized, ultra-weakly
continuous, functionals on $\ga.$ They are represented by
positive, trace class operators $\rho,\, \mbox{Tr}\, (\rho )=1,$
with $\rho (A)\doteq\mbox{Tr}\, (\rho A),$ $A\in \ga .$  Pure
states are represented by $\rho$ of the form $\rho=P$, where $P$
is an orthogonal projection onto a 1-dimensional subspace of
$\gH$. The space of pure states can be thus identified with the
space of one-dimensional subspaces of $\gH.$ If $\gH$ is
finite-dimensional, $\gH\approx\Cset^n$, then the space of pure
states is the complex projective plane $CP^{n-1}.$ Quantum
dynamics is usually described in terms of a 1-parameter group of
unitary operators $U(t):t\in\bR.$ It acts on observables via
automorphism $\alpha_t: A\mapsto U(t)^{-1}AU(t).$
\subsection{Mixed quanto-classical dynamics} We will consider the
simple case, where the classical system is finite $X=\{1,\ldots
,N\}.$ For each $\alpha\in X$ consider the Hilbert space
$\gHa=\Cset^{n_\alpha}$ and let $M(n_\alpha)$ be the algebra of
$n_\alpha\times n_\alpha$ complex matrices.\footnote{In all
examples studied so far the dimensions $n_\alpha$ were the same
for all $\alpha.$ But such a restriction is not necessary.} The
observables of the coupled system are now functions $\alpha\mapsto
A_\alpha\in M(n_\alpha)$ on $X$ with values in $M(n_\alpha)$. A
pure state of the system is a pair $(\alpha,P)$, where
$\alpha\in\{1,\ldots ,N\}$ and $P$ is a Hermitian projection
matrix onto a one-dimensional subspace in $\Cset^{n_\alpha}.$ It
is not possible to couple the classical and the quantum degrees of
freedom via reversible, unitary dynamics. A 1-parameter semi-group
of completely positive maps of the algebra
$\ga=\oplus_{\alpha=1}^N M(n_\alpha)$ is being used instead. We
are interested in semi-groups with generators of Lindblad's type
(also known as ``dynamical semigroups"), in particular with
generators of the form: \be
L(A)_\alpha=\mbox{i}[\ha,\aaa]+\sum_{\beta\neq\alpha} \gba^\star
A_\beta\ \gba - \frac{1}{2}(\la\aaa+\aaa\la ),\label{eq:lioua} \ee
where $\gba \in L({\go H}_\alpha,{\go H}_\beta)$ and \be
\la=\sum_{\beta\neq\alpha} \gba^\star \gba \in L(\gHa ).\ee We
always assume that the diagonal terms vanish:
$g_{\alpha\alpha}=0.$ It has been shown in \cite{jakol95} that
there is a one--to--one correspondence between semigroups with
generators of the above type and piecewise-deterministic Markov
processes on the space of pure states of the system.
\subsection{From dynamical semigroups to QIFS\label{sec:ds}} Here
we are not concerned with the continuous time evolution between
jumps, so let us extract from the Ref. \cite{jakol95}, and also
slightly reformulate, the jump process alone. It is determined by
the operators $\gab$ alone, and it is an iterated function system,
with place-dependent probabilities that are also determined by
$\gab$-s. Let $(\alpha,P)$  be pure state, with $P$ being an
orthogonal projection on a unit length vector $\psi\in\gHa.$
Observe that for each $\beta\neq\alpha$ we have: \be \gba P
\gba^\star = \lambda(\alpha,\beta;P)Q\label{eq:pgba} \ee where \be
\lambda(\alpha,\beta;P)=\Vert \gba\psi\Vert^2\geq0 \ee and, if
$\lambda(\alpha,\beta;P)>0,$ then $Q$ is a projection operator on
the vector\newline $\gba\psi/\Vert \gba\psi\Vert$ in $\gH_\beta.$
The probabilities $p(\alpha,\beta;P)$ are defined as \be
p(\alpha,\beta;P)=\frac{\lambda(\alpha,\beta;P)}{\sum_{\beta\neq\alpha}\lambda(\alpha,\beta;P)}.\ee
Assume now that all Hilbert spaces $\gHa \equiv\gH$ are identical.
Assume that $X=2^N$ - the set of $N$ bits, and that $\gab=g_i\neq
0$ when $\alpha$ differs from $\beta$ only at one, the $i$-th bit,
otherwise $\gab=0.$ We will just have a family of operators $g_i$
and a jump process on pure states $P$, that is one-dimensional
orthogonal projections in $\gH.$ The maps and their probabilities
are determined by: \be g_i P g_i^\star = \lambda(i;P)Q,\ee with
$\lambda(i;P)=\Vert g_i\psi\Vert^2,$ $p_i(P)= \lambda(i;P)/\sum_j
\lambda(j,P)$, and $Q$ being the orthogonal projection on the
subspace spanned by the vector $g_i\psi.$ We have an iterated
function system on the complex projective space $CP(n-1)$
(equivalently, on the grassmannian of one-dimensional subspaces of
$\Cset^n$), with place dependent probabilities. Both, maps and
probabilities, are determined by the set of linear operators
$g_i,\quad i\in \{1,2,\ldots ,N\} $ \subsection{A short history of
QIFS} The idea of coupling a classical and a quantum system via
dynamical semigroup has been originally described in
\cite{blaja93a}. The first model of a QIFS, on $S^2$, with index
$\alpha$ being also continuous $\alpha=\n$, with values in $S^2$
and, using the notation of Sec. \label{seq:alg} $g_\n$ defined as
$g_\n=\exp\left( i\pi\sigma(\n )\right).$ These maps were unitary,
thus measure preserving, and did not give rise to a fractal
attractor. The tetrahedral model was first introduced in
\cite{jadifs}. It was then examined analytically and modelled
numerically in a PhD Thesis by G. Jastrzebski \cite{gjast}. The
model was further exploited in \cite{blajaol99a}, where it has
been described in some details, and where the Lyapunov exponent of
the semigroup generator has been computed. The term QIFS has been
introduced about that time on sci.physics.research newsgroup on
internet. Recently the term QIFS has been adopted in \cite{loslzy}
for another class of maps, namely for maps on the space of all
statistical states of a quantum system, with arbitrarily assigned
probabilities. \subsection{Some open problems} The question of
uniqueness of the invariant measure for a general QIFS is an open
technical problem. For a particular case of quantum octahedron we
used numerical estimations of the average contraction parameter.
We do not have an analytical proof even in this particular case.
Then there is a "philosophical" question: can the fractal patterns
derived from the QIFS algorithm be "observed"? Or are they purely
mathematical constructs that have no relation to the "real world"?
The question is not an easy one to answer, because QIFS live in
the projective Hilbert space of pure states of a quantum system,
not in "our space". The question of whether they can be observed
is related to the question: can the wave function be observed? A
preliminary discussion of this problem has been given in Ref.
\cite{jad94a}. We hope to return to this question in our future
publications.

\begin{thebibliography}{99} \bibitem{barnsley}
M.~F.~Barnsley, {\it Fractals everywhere}, (Academic Press,  San
Diego 1988) \bibitem{skala} L.~Skala, K.~Bradler and V.~Kapsa,
\newblock Consistency requirement and operators in quantum
mechanics, {\it Czech. J. Phys.} {\bf 52} (2002), 345-350.
\bibitem{jadob02}A.~Jadczyk and R.~{\"O}berg, \newblock Quantum
Jumps, EEQT and the Five Platonic Fractals, \newblock Preprint:
http://arXiv.org/abs/quant-ph/0204056
\bibitem{gjast}G.~Jastrzebski, \newblock Interacting classical and
quantum systems. Chaos from quantum measurements, \newblock Ph.D.
thesis, University of Wroc{\l}aw, 1996 (in Polish)
\bibitem{stenflo02}{\"O}.~Stenflo, \newblock Uniqueness of
invariant measures for place-dependent random iterations of
functions, \newblock Published in IMA Vol. Math. Appl., 132
(2002), 13--32, \newblock Preprint:
http://www.math.su.se/~stenflo/IMA.pdf \bibitem{barnsley88}
M.~F.~Barnsley, S.~G.~Demko, J.~H.~Elton and J.~S.~Geronimo,
Invariant measures for Markov processes arising from iterated
function systems with place-dependent probabilities, {\it Ann.
Inst. H. Poincar{\'e} Probab. Statist} {\bf 24} (1988) 367-294,
(Erratum: {\bf 25} (1989) 589--590 \bibitem{jakol95} A.~Jadczyk,
G.~Kondrat and R.~Olkiewicz, \newblock On uniqueness of the jump
process in quantum measurement theory, \newblock {\it J. Phys.
A}\, {\bf 30} (1996) 1-18; Preprint
http://arXiv.org/abs/quant-ph/9512002 \bibitem{blaja93a}
Ph.~Blanchard and A.~Jadczyk, On the Interaction Between Classical
and Quantum Systems {\it Phys. Lett. A} {\bf 175} (1993), 157-164;
Preprint http://arXiv.org/abs/quant-ph/9512002 \bibitem{jad94a}
A.~Jadczyk, Topics in Quantum Dynamics, in Proc.~First Caribb.
School of Math.~and Theor.~Phys., Saint--Francois--Guadeloupe
1993, {\it Infinite Dimensional Geometry, Noncommutative Geometry,
Operator Algebras and Fundamental Interactions}, ed. R.Coquereaux
et al., (World Scientific, Singapore 1995); Preprint
http://arXiv.org/abs/hep-th/9406204 \bibitem{jadifs} Jadczyk, A.:
\lq IFS Signatures of Quantum States\rq, IFT Uni Wroclaw, internal
report, September 1993. \bibitem{blajaol99a} Ph.~Blanchard,
A.~Jadczyk and R.~Olkiewicz, \newblock Completely Mixing Quantum
Open Systems and Quantum Fractals, \newblock {\it Physica D:
Nonlinear Phenomena}\ , {\bf 148} (3-4) (2001) 227--241; Preprint
http://arXiv.org/abs/quant-ph/9909085 \bibitem{loslzy}
A.~Lozinski, K.~Zyczkowski and W.~Slomczynski, \newblock Quantum
Iterated Function Systems, \newblock  Phys. Rev. E  {\bf 68}
(2003), 046110; Preprint http://arXiv.org/abs/quant-ph/0210029
\end{thebibliography}
 \end{document}